\documentclass[fleqn,twoside]{article}
\usepackage{espcrc2}

\usepackage{graphicx}

\newcommand{\AmS}{{\protect\the\textfont2
  A\kern-.1667em\lower.5ex\hbox{M}\kern-.125emS}}

\usepackage{natbib,psfig}

\def\HI{H{\,\small I}}
\newcommand{\kms}{$\,$km$\,$s$^{-1}$}
\newcommand{\mJybeam}{mJy beam$^{-1}$}

\newcommand{\msun}{{$M_\odot$}}

\title{Disks, Tori, and Cocoons: Emission and Absorption Diagnostics of AGN
Environments
}
\author{R. Morganti
\address{Netherlands Foundation for Research in Astronomy, \\
Postbus 2, NL-7990 AA, Dwingeloo, NL; morganti@astron.nl},
L. J. Greenhill\thanks{Current address: Visiting physicist, Kavli Institute for Particle Astrophysics
and Cosmology, Stanford University.}
\address{Harvard-Smithsonian Center for Astrophysics, 60 Garden St, Cambridge,
MA 02138 USA; greenhill@cfa.harvard.edu},
A.~B. Peck\address{Harvard-Smithsonian Center for Astrophysics, SMA Project, 645 N. 
A'ohoku Pl., Hilo, HI 96720, USA; apeck@cfa.harvard.edu},
D.~L. Jones\address{Jet Propulsion Laboratory,
    California Institute of Technology, 4800 Oak Grove Drive,
    Pasadena, CA 91109 USA; dj@sgra.jpl.nasa.gov},
C. Henkel\address{MPIfR, Auf dem Huegel 69, D-53121 Bonn, Germany;
p220hen@mpifr-bonn.mpg.de}}

\begin{document}

\begin{abstract}

One of the most important problems in the study of active galaxies is
understanding the detailed geometry, physics, and evolution of the central
engines and their environments.  The leading models involve an accretion disk
and torus structure around a central dense object, thought to be a
supermassive black hole.  Gas found in the environment of AGN is associated
with different structures: molecular accretion disks, larger scale atomic
tori, ionized and neutral ``cocoons" in which the nuclear regions can be
embedded. All of them can be studied at radio wavelengths by various means.
Here, we summarize the work that has been done to date in the radio band to
characterize these structures.  Much has been learned about the central few
parsecs of AGN in the last few decades with contemporary instruments but the
picture remains incomplete. In order to be able to define a more accurate
model of this region, significant advances in sensitivity, spectral and
angular resolution, and bandpass stability are required.  The necessary
advances will only be provided by the Square Kilometer Array and we discuss
the possibilities that these dramatic improvements will open for the study of
the gas in the central region of AGN.

\end{abstract}

\maketitle





\section{Introduction}

The physical conditions in active galactic nuclei (AGN) are unique in the
cosmos.  Stellar and gas densities are very large, and enormous amounts of
angular momentum and energy are released as material accretes onto massive
black holes.  Stellar and interstellar gas constitute reservoirs of accreting
material.  Study of the structure, kinematics, and excitation of this material
is the sole means available to directly study massive compact objects, which
are not otherwise directly visible.  While stars close to massive black holes
are difficult to detect because of obscuration and crowding, emission and
absorption by the neutral atomic, ionized, and molecular components of the
interstellar medium (ISM) may be readily studied at radio wavelengths.

The role of the ISM in active nuclei is an important one because it feeds the
central engines, thus determining their masses and angular momenta.  In
addition, material directly affects the overall appearance of AGN in two ways.
First, it affects the degree of shielding of the central engines from various
viewing angles, recognition of which motivated formulation of the AGN
unification paradigm.  Second, the accreting gas emits intense electromagnetic
radiation from radio to X-ray bands, which affects the ISM structure and
energetics in parts of the parent galaxies and provides a handle for the study
of matter under truly extreme conditions.

Gas structures (from disks to tori to ``cocoons'') in AGN have been
extensively studied at radio wavelengths, providing a wealth of important
results that have been crucial to building our present picture of AGN.
However, the picture remains incomplete because the studies haved been limited
by the sensitivity, spectral resolution, instantaneous bandwidth, and bandpass
stability of present and past instruments.  We discuss in detail the
substantially deeper studies of AGN that will be possible through the dramatic
improvements that can be obtained with observers with the SKA.

\section{Accretion disks and tori}

The AGN unification paradigm (Antonucci 1993) posits that central engines are
surrounded by a parsec-scale toroidal (or disk-like) distribution of atomic or
molecular gas that obscures lines of sight for which the radio axis is close
to the plane of the sky.  Some fraction of the gas accretes to the central
engine and is responsible for observed high luminosities.  Tori can exist
where energetic processes can provide vertical support, e.g. as in heating or
magnetic turbulence at sub-parsec radii.  In cases where infalling material
can cool, it forms a thin accretion disk (Shakura \& Sunyaev 1973), which may
be warped by radiative, magnetic, gravitational, and relativistic effects
(Pringle 1996, Neufeld \& Maloney 1995).  For example, radiative warping can
be driven by X-ray heating of a disk and anisotropic reradiation of the energy
by disk material.  As is true for tori, warps shadow material at larger radii
and similarly affect the appearance of the AGN for different observers.

The most challenging AGN to study are those whose radio axes are close to the
plane of the sky, because of high obscuration (at optical and infrared
wavelengths) and a superposition of structures along the line of sight.  At
radio wavelengths, signatures of emission and absorption by molecular gas,
absorption by atomic gas, and free-free absorption of continuum emission are
the most commonly used probes.

In the radio, there are three methods by which it is possible to
detect circumnuclear structure in objects whose jet axis lie close to
the plane of the sky: molecular gas seen either masing or in
absorption, atomic gas seen in absorption, and ionized gas revealed
through free-free absorption.

\begin{figure*}
\vskip 3cm
\centerline{ Here Fig1.tif}
\vskip 3cm
\caption{Representations of warped disk models for two AGN.  
Color codes mark the Doppler shift.  Light smudges indicate the mapped
positions of maser emitting regions.  In NGC\,4258, the continuum
emission from a narrowly collimated jet has been observed in line-free
interferometer channels and registered precisely with respect to the
dynamical centre of the disk (Herrnstein et al. 1997).  Circinus is
not known to have a jet, but the edges of a well known
kiloparsec-scale outflow and ionization cone correspond well to the
edges of the shadow created by the disk warp.  ``Off-disk'' masers are
also seen in the unshadowed region but at sub-parsec radii, where they
likely trace a clumpy, high density flow that arises possibly
$<0.1$~pc from the central engine.}
\end{figure*}

\subsection{Mapping H$_2$O masers $<$ 1pc from massive black holes}

Over 50 type-2 active galactic nuclei contain known sources of H$_2$O maser
emission ($\nu_{\rm rest}\sim22$~GHz), a large fraction discovered since 2000
(see Figure 3, Greenhill this Volume).
Radio interferometric studies of these ``nuclear masers'' are the only means
by which structures $<1$ pc surrounding supermassive black holes can be mapped
directly, which is particularly important for type-2 nuclei because edge-on
orientation and obscuration complicate the study by other means.
Investigations of several sources have demonstrated that H$_2$O maser emission
traces warped accretion disks 0.1 to 1~pc from central engines with masses of
order $10^6$-$10^8$~M$_\odot$.  Though not yet mapped, almost half the known
nuclear masers share spectral characteristics consistent with emission from
edge-on accretion disks.  Their mapping and the discovery of new nuclear
masers are high priorities.  Instrument requirements include sensitivity to
sub-mJy lines, instantaneous GHz bandwidths, spectral resolutions on the order
of tens of kHz, and sub-milliarcsecond resolution for imaging.  
Built to current specifications, the SKA will provide a {\it two} order of
magnitude increase in spectroscopic sensitivity for surveys intended to detect
new maser sources in AGN.  This should permit discovery of new sources at  
cosmological redshifts.  The Array will enable a {\it one} order of magnitude
improvement in sensitivity for (follow-up) high angular resolution imaging,
where the SKA must be combined with an array of outrigger antennas operating
on intercontinental (and ground-space) baselines (see Section 4.1).

\begin{figure*}
\centerline{\psfig{figure=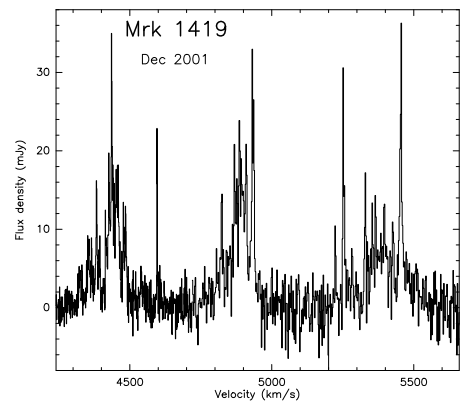,angle=-90,width=8cm}}
\caption{22\,GHz H$_2$O spectrum of Mrk\,1419 (NGC\,2960), showing the characteristic
systemic (center), red- and blue-shifted groups of maser features that
are hinting at an edge-on accretion disk.}
\end{figure*}

\subsubsection{What are nuclear H$_2$O masers?}

Water maser emission from galactic nuclei is a beacon of accretion
onto supermassive black holes.  Recent general reviews may be found in
Maloney (2002), Greenhill (2002, 2004), Watson (2002) and Henkel et
al. (2004).  On the order of $10^3$ galaxies
($cz<2\times10^4$~km\,s$^{-1}$) have been observed in searches for new
maser sources, with a variety of sensitivities and bandwidths.
Somewhat more than 50 have been detected (see Figure 3 in Greenhill, this
Volume). The phenomenon is associated generally with Seyfert-2
galaxies and LINERS (Braatz et al. 1996), though there are examples in
narrow line Seyfert-1 systems (Hagiwara et al. 2003), transition
objects between Seyfert 1 and 2 classes (e.g., Braatz et al. 1996),
and FR-II radio galaxies (Tarchi et al. 2003).  The spectra of almost
half the known masers exhibit emission over a broad range of velocity
that (more or less) symmetrically brackets the systemic velocities of
the host galaxies.  The emission is distributed in plateaus from one
velocity extreme to the other or in distinct line complexes, at large
offset velocities and near the systemic velocities of the host
galaxies.

Long baseline interferometers have resolved the angular and velocity
structure of several maser sources each with emission spread over up
to $\sim 2200$~km\,s$^{-1}$ ($\sim 160$~MHz).  This has demonstrated
that the emission originates in nearly edge-on accretion disks.  In
addition to position and Doppler velocity, proper motions and secular
drifts in line-of-sight velocities have also been measured for
material near systemic, which correspond to the rotation and
centripetal acceleration of the disk orbits, respectively.  The best
established cases are NGC\,4258 (Humphreys et al. 2004, in prep;
Herrnstein et al. 1999; Moran, Greenhill, \& Herrnstein 1999, and
references therein), NGC\,1068 (Greenhill \& Gwinn 1997), and the
Circinus galaxy (Greenhill et al.  2003). In NGC~4258, Herrnstein et
al (1997) have used the VLBA to determine the size and shape of the
warped molecular disk as traced by maser spots.  In this source, the
masing disk extends from 0.13 pc to 0.25 pc from the central engine
(see Fig.~1).  Maps of maser emission in NGC\,3079 (Trotter et
al. 1998; Kondratko, Greenhill, \& Moran 2004), are also suggestive of
emission from a disk, while evidence for NGC\,4945 (Greenhill et
al. 1997), IC\,2560 (Ishihara et al. 2001), NGC\,5793 (Hagiwara et
al. 1997), IRAS\,F22265-1826 (Ball et al. in
prep.), and NGC\,1386 (Greenhill \& Braatz, unpublished), is less
complete.

In the firmly established cases, NGC\,4258, NGC\,1068, and Circinus, maser
emission traces well ordered disk material at radii of 0.1-1~pc and rotation
speeds up to $\sim 1100$~km\,s$^{-1}$ around central engines of
$10^6$-$10^8$~M$_\odot$.  Disk shape, orientation with respect to the line of
sight, velocity shear, and background amplification largely determine where
favoured maser gain paths are found within the disks (i.e., chiefly where the
orbital motion is nearly parallel or perpendicular to the line of sight).  To
first order, the high perceived brightness of nuclear masers, compared to
masers in star forming regions, can be accounted for by the long amplification
paths made possible by the large-scale orderly dynamics of accretion disks,
which (anisotropically) beam maser radiation in planes tangent to disk
surfaces. However, pumping considerations are also important.  The most
broadly applicable mechanism is probably heating of disk gas via oblique
irradiation by a central X-ray power-law source (Neufeld \& Maloney 1995).  It
is disk shape (e.g., warping) that determines which surfaces and regions are
irradiated and which are not (i.e., shadowed).  In addition to the three
established cases, masers in NGC~1386 also probably trace a warped accretion
disk $\sim 1$~pc across around a central mass on the order of
$10^6$~M$_\odot$.  The inferred central mass for disks in NGC\,3079 (Trotter
et al. 1998, Kondratko et al. 2004) and IRAS\,F22265-1826 (Ball et al. 2004)
are similar, but these disks are relatively large, 1-3~pc in radius, and not
well ordered kinematically, probably because they are gravitationally unstable
to fragmentation and star formation (e.g., Goodman 2003; Levin \& Beloborodov
2003; Milosavljevi\'c \& Loeb 2004).  In the cases of IC\,2560 (Ishihara et
al. 2001) and Mrk\,1419 (Henkel et al. 2002; see Fig.\,2), full disk models
are not yet available, but central masses of $3\times10^6$ and $\sim
10^7$~M$_\odot$, respectively, have been obtained using reported accelerations
and estimates of orbital velocity.

\subsubsection{Key science with  nuclear H$_2$O masers}

As already mentioned, interferometric observation of masers is presently
the only means by which structures $<1$~pc from supermassive black holes can
be mapped.  This is especially important for type-2 systems because in them,
the central parsec is viewed close to edge-on, such that obscuration is heavy.
Reverberation mapping of broad line emission is impractical and modeling of
spectroscopic data (e.g., X-ray) must account for the superposition of
multiple emission and absorption components along the line of sight.  Examples
of key science unique to maser studies follow.

{\it Sub-parsec dynamical masses for central engines---} Mass
estimates enable calculation of Eddington luminosities and accretion
efficiencies.  For disks whose rotation curves can be resolved and a
simple geometric model fit, the fractional uncertainty in central mass
is dominated by the fractional uncertainty in distance to the host
galaxy.  Ultimately, maser dynamical masses for a large sample of
galaxies will establish independently the slope of the
$M_\bullet-\sigma$ relation (e.g., Ferrarese \& Merritt 2000) and
intrinsic scatter therein, limited chiefly by uncertainties in the
stellar velocity dispersions in nuclei.

{\it The 3-D shapes of accretion disks---} Disks that have been mapped provide
strong evidence that accretion disks can be warped by 0.1-1 radian on 0.1-1 pc
scales.  However, the warping mechanism is not certain. Though radiative
torques have been suggested (Neufeld \& Maloney 1995, Pringle 1996),
stability is a concern.  Testing of warp models requires characterisation of
warps for a sample of AGN that contain masers and exhibit a range of
luminosities and accretion rates.

{\it Accretion disk thickness---} Estimates of thickness are critical
to calculations of accretion rate and identifications of accretion
modes (e.g., advective, convective, viscous).  No thicknesses have
been measured directly, but high angular resolution observations of
masers provide perhaps the best chance.  Limits established for
NGC\,4258 are the tightest so far, and they are consistent with
hydrostatic equilibrium (Moran et al. 1999).  However, because
NGC\,4258 is underluminous, this disk thickness may not be typical.

{\it Heterogeneity among type~2 AGN---} Because of the high gas
density required to support maser action, disks that cross the lines
of sight to central engines are ready substitutes for the obscuring
tori featured in the AGN unification paradigm.  These tori have been
difficult to observe directly and in large numbers.  In contrast,
warped accretion disks (at least in nuclei that host masers) appear to
be common and their shapes suggestive. Outflows close to central
engines may also provide the observed absorption columns (e.g.,
Fig.~1).

\begin{figure*}
\vskip 3cm
\centerline{ Here Fig3.gif}
\vskip 3cm
\caption{{\sl (Left)} The absorption profiles toward each of the 6 resolved
continuum components across 1946+708.  The systemic velocity is
indicated by an arrow in each profile.  The velocity resolution is 16
\kms.  The beam, shown in the lower left corner, is
4.3$\times$4.9 mas.  The linear scale shown in the lower right assumes
H$_0$=75 \kms\ Mpc$^{-1}$.
{\sl (Right)} A cartoon of how a circumnuclear torus might look.  Some
notable simplifications have been made.  For example, it is not
necessary for the toroidal structure to be perpendicular to the jet
axis, the ``clumps'' of denser gas are unlikely to be uniform in size,
and the degree of warp in the disk can vary greatly.
Also, the relative scale heights and radii of the various regions are
not yet well constrained.  This model represents only the central
$\le$100 pc of the source, and does not address the possibility of an
extended \HI\ or molecular disk or torus associated with the host
galaxy, although this could conceivably be part of the same continuous
structure.  The extent of the inner molecular disk is likely to be on
the order of a parsec, and the region of atomic gas beyond that
probably extends some 50-100 pc from the central engine.  This outer
region, although predominantly atomic, probably contains clumps or
clouds of denser gas which could harbour molecular gas.}
\end{figure*}

\subsection{Gas in the circumnuclear tori}

\subsubsection{The neutral hydrogen component}

Evidence of circumnuclear \HI\ is detectable in absorption in radio sources
whose jet axes lie close to the plane of the sky (Conway 1997), particularly
in young compact sources ($\sim$10$^4$~yr, Readhead et al 1996, Vermeulen
et al. 2003).  Evidence of
a circumnuclear torus of atomic gas has been seen in Cygnus A (Conway 1999),
NGC 4151 (Mundell et al 1995) and 1946+708 (Peck et al. 1999).  In Cygnus A,
\HI\ absorption measurements with the VLBA indicate a torus with a radius of
$\sim$ 50 pc.  In NGC 4151, \HI\ absorption measurements using MERLIN
indicate a torus $\sim$ 70 pc in radius and $\sim$ 50 pc in height.
One of the best examples of this type of torus is the Compact
Symmetric Object (CSO) 1946+708.  The \HI\ absorption in 1946+708
consists of a very broad line and a lower velocity narrow line which
are visible toward the entire $\sim$100 pc of the continuum source
(see Fig.~3 (left), Peck et al. 1999). The broad line has low optical
depth and peaks in column density near the core of the source.  This
is consistent with a thick torus scenario in which gas closer to the
central engine is much hotter, both in terms of kinetic temperature
and spin temperature, so a longer path-length through the torus toward
the core would not necessarily result in a higher optical
depth. Figure 3 (right) shows a cartoon of what this structure might
look like. The high velocity dispersion toward the core of 1946+708 is
indicative of fast moving circumnuclear gas, perhaps in a rotating
toroidal structure.  Further evidence for this region of high kinetic
energy and column density is found in the spectral index distribution
which indicates a region of free-free absorption along the line of
sight toward the core and inner receding jet.  The \HI\ optical depth
increases gradually toward the receding jet.  The most likely scenario
to explain these phenomena consists of an ionized region around the
central engine, as well as an accretion disk or torus, with a scale
height of $\le$10 pc at the inner radius and at least 80 pc at the
outer radius, which is comprised primarily of atomic gas (Peck \&
Taylor 2001).

Due to the weakness of the radio core in many sources, the study of the
\HI\ associated with circumnuclear tori can be done only for a limited number
of cases, namely those with compact jets. The extension of this study to radio
sources of different power and ages is crucial.  For example, in low
luminosity radio galaxies the situation could be different.  In these sources,
the standard pc-scale geometrically thick torus is perhaps not present (as
obtained from optical and X-ray studies). Indeed, the presence of a thin disk
has been determined from \HI\ observations in the case of NGC~4261 (van
Langevelde et al.\ 2000).  For this object, the VLBI data suggest that the
\HI\ absorption is due to a disk $\sim$1.3 pc thick seen in projection against
the counter-jet.  Differences in the nuclear torus/disk system (and
corresponding difference in the accretion processes) is key to understanding
the observed variations in radio power, radio morphology, optical ionisation,
and other parameters.

The sensitivity of the SKA will allow radio counterjets to be detected in a
much larger sample of sources spanning a wide range of luminosities and
orientations.  Thus, the SKA will reduce the observational bias in future studies
of source symmetry and absorption probes of the circumnuclear region.  As a
result, correlations between the gas properties and geometry in the inner
parsecs of active nuclei and other properties of the sources will be easier to
measure and understand.

\begin{figure*}
\vskip 3cm
\centerline{ Here Fig4.gif}
\vskip 3cm
\caption{{\sl (Left)} VLBA  image at 4.9~GHz of the nucleus of
NGC 4261 (Jones et al.  2001), showing a radio jet and counterjet and a deep
and narrow gap in emission just on the counterjet side
of the core. {\sl (Right)} The distribution of
radio spectral index (between 4.9 and 8.4~GHz) across the central few parsecs of
NGC 4261 (Jones et al.  2001).  Note that the region of minimum emission just
east of the brightest peak has a strongly inverted
(absorbed) spectrum, while the peak (core) has a less
inverted spectrum and the jet and counterjet have flat
to steep spectra.
}
\end{figure*}

\begin{figure*}
\vskip 3cm
\centerline{ Here Fig5.gif}
\vskip 3cm
\caption{
{\sl (Left)} Two-dimensional distribution of the free-free absorption in 3C 84
over the region of the counterfeature (from Walker et al. 2000). There is no
detected absorption over the brighter southern feature. The overlaid contours
are from the 8.4 GHz image while the spectral index is determined from a fit
to data at four frequencies (5.0, 8.4, 15.4, and 22 GHz). The strong radial
gradient away from the core is apparent.  {\sl (Right)} Montage of the VLBA
images of 3C 84 (from Walker et al. 2000). These displays are based on the
CLEAN components convolved with a common Gaussian beam of 1.6 $\times$ 1.2
mas, elongated north-south. The contour levels start with 5, 10, 14, 20 mJy
beam$^{-1}$ and increase from there by factors of $\sqrt 2$.  The north-south
segmented line shows the location of the slice along which some of the
analysis was done. Note that, for H$_0$ = 75 km s$^{-1}$ Mpc$^{-1}$, the scale
is about 3 mas pc$^{-1}$.}
\end{figure*}

\subsubsection{Maser emission and the circumnuclear tori}

Once evidence for a torus has been found in \HI, these sources are prime
candidates to search for molecular absorption, maser emission and ionized gas
in the central parsecs.  Observations of NGC 1068, for example, suggest
evidence of all three components, consisting of OH and H$_2$O masers, \HI\
absorption, and a compact source of free-free emission (Gallimore et al 1996).
As mentioned above, in this source, the maser emission indicates that the
molecular disk makes a large angle with the radio axis (Greenhill \& Gwinn
1997) and so might be more strongly warped than that of
NGC4258. IRAS\,F22265-1826 contains the most luminous known H$_2$O megamaser,
with a luminosity in the 1.3 cm line of 6100 L$_{\odot}$; (Koekemoer et
al. 1995).  Although the ~420 km s$^{-1}$ absorption probably results from
neutral material associated with the atomic and molecular torus thought to
feed the active nucleus, the deeper ~125 km s$^{-1}$ line in TXS 2226-184
could be indicative of an interaction between the radio jet and surrounding
material (Taylor et al. 2002). The main component of the water maser emission
is also fairly broad (almost 100\,km\,s$^{-1}$) in TXS 2226-184 and could
likewise originate from the central torus or from a jet-cloud interaction.

To date, however, too few sources have been found which exhibit both
\HI\ absorption and H$_2$O maser emission to confirm this scenario.
Of the 19 H$_2$O megamaser galaxies presented in Taylor el al.
(2002), 7 are known to have associated \HI\ absorption. Likewise, 4
\HI\ torus systems (1946+708: Peck, Taylor, \& Conway 1999; PKS 2322-123:
Taylor et al. 1999; Hydra A: Taylor 1996, NGC 3894: Peck \& Taylor 1998) have
been searched for maser emission with no detections.  A fifth \HI\ torus
source (NGC 4151: Mundell et al. 1995) has recently been found to exhibit very
weak maser lines (Braatz et al. in prep). Part of the reason for this may be
orientation effects, insofar as the maser emission is more heavily dependent
on the line of sight through, and the degree of warp in, the accretion disk
itself.  Much of the difficulty, however, might simply stem from a lack of
sensitivity.  Many megamaser sources have very little 21cm continuum emission
toward which to detect the HI absorption.  In addition, increasingly faint
masers are being detected with every improvement to the 22 GHz receiver
systems at existing telescopes, to the point that detections of extragalactic
megamasers with peak flux densities of 10-20 mJy are becoming fairly common
(Peck et al. in prep.).  In both cases, building a statistical sample relies
solely on increased sensitivity and moderate angular resolution, such as would
be achieved with the SKA.

\subsubsection{Free-Free absorption by accretion disks }

Parsec-scale radio counterjets are important for studying the 
intrinsic symmetry of the jet-formation process, and as probes of 
the structure of ionized gas in the central pc of accretion disks 
surrounding the central black holes in AGN.  In this latter case, the 
counterjet serves as a source of radiation which can be absorbed by 
thermal electrons in the disk.  If geometrically thin, the disk will 
cover the inner part of the counterjet but not the approaching jet.
If geometrically thick, the core and perhaps the base of the 
approaching jet may also be absorbed, but with lower total optical 
depth than the base of the counterjet.

Multi-frequency observations can detect the highly inverted spectrum created
by free-free absorption from flat or steep spectrum synchrotron emission from
the radio core and jet. Two well studied cases are shown in Figs. 4 and
5. Even in cases where the radio core has an inverted spectrum because of
synchrotron self-absorption, it is possible to use the differing angular size
of free-free absorbing regions as a function of frequency to distinguish this
process from synchrotron self-absorption.

The SKA will allow such studies of disks in a much larger sample of 
objects, covering a wider range of jet and disk orientations, and 
will detect absorption over a larger range of disk radii.  By 
measuring the free-free optical depth as a function of projected disk 
radius, and knowing the disk orientation from the radio jet observations 
(comparisons of jet and counterjet brightness and proper motions), we 
can solve for the radial distribution of plasma density in the disk.

In addition, the SKA will be able to routinely add Faraday rotation 
and depolarisation constraints (from multi-frequency polarisation 
measurements) to models of the electron density and magnetic field 
distributions with the accretion disks.  The radial distribution of 
free electrons in the disk can be used to estimate the mass accretion 
rate $dm/dt$ (Kungic \& Bicknell, in prep.); this in turn can be 
compared with the observed jet energetics to provide critical input 
parameters for all models of the central engine.

Accretion disks are a fundamental component of the central engines that
produce relativistic jets in objects from galactic microquasars containing
stellar mass black holes to powerful AGN containing billion solar mass black
holes.  It is very likely that
relativistic jets cannot form without an accretion disk.  Consequently the
study of the basic physics of central engines requires knowledge of the
physical properties of these disks.  The SKA will revolutionise our ability to
determine many of the relevant physical parameters.


\begin{figure}
\centerline{\psfig{figure=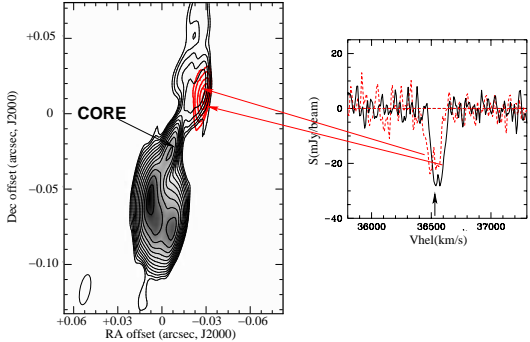,angle=0,width=8cm}}
\caption{Continuum image (grey scale and thin contours) of 4C~12.50 (from
Morganti et al. 2004a)
superimposed onto the total intensity of the \HI\ absorption (thick
contours). The position of the radio core  is also indicated. The   
contour levels for the continuum image are: 5 \mJybeam\ to 800 \mJybeam\ in
steps of factor 1.5.}
\end{figure}

\section{Rich ISM and the evolution of the radio sources}

The origin of activity
 in galaxies is often explained as
having been triggered by merger and/or interaction processes.  Torques and
shocks during the merger can remove angular momentum from the gas in the
merging galaxies and this provides injection of substantial amounts of
gas/dust into the central nuclear regions (see e.g.  Mihos \& Hernquist 1996). 
It is, therefore, likely that in the initial phase of an AGN, this gas still
surrounds (and possibly obscures) the central regions.  AGN-driven outflows
have major effects on this dense ISM.  This feedback process can also play an
important part in the evolution of the host galaxy as it may limit the growth
of the supermassive black holes (BHs) and regulate the correlation between BH
and bulge properties (see e.g.  Silk \& Rees 1998).

\HI\ absorption can be used to trace such a rich medium.
One example is the radio galaxy 4C~12.50 (see Fig. 6, Morganti et al. 2004a), a
galaxy that has often been suggested to be a prime candidate for the link
between ultraluminous infrared galaxies and young radio galaxies. In this
object, deep and relatively narrow \HI\ absorption (observed at the systemic
velocity) is associated with an off-nuclear cloud ($\sim 50$ to 100 pc from
the radio core) with a column density of $\sim 10^{22}\ T_{\rm spin}/(100\
{\rm K}$) cm$^{-2}$ and an \HI\ mass of a few times $10^5$ to $10^6$
\msun. There are more examples of objects where the \HI\ traces the rich
medium surrounding the active nucleus. Examples of off-nuclear \HI\ absorption
are found in 3C~236 (Conway \& Schilizzi 2000) and, more recently,  in
the CSO 4C~37.11 (Maness et al.\ 2004) where a broad
($\sim 500$ \kms) absorption line was found in the region of the southern
hot-spot.

The relevance of this is that it may have important implications
for the evolution of the radio jets. Although this gas will not be able
to confine the radio source, it may be able to momentarily
destroy the path of the jet as shown also by numerical simulations
(Bicknell et al.\ 2003). Thus, this interaction can influence the
growth of the radio source until the radio plasma clears its way out.

\subsection{Outflows of neutral hydrogen}

The scenario described above, where the central region could be still
surrounded by a rich cocoon of ISM, implies that gas outflows can be generated
by the nuclear engine interacting with such a medium.  The discovery of a
number of radio galaxies where the presence of fast ($> 1000$ \kms) outflows
is associated with {\sl neutral hydrogen} is thus extremely intriguing.  This
finding gives new and important insights into the physical conditions of the
gaseous medium around an AGN.  The best examples so far are the radio galaxies
3C~293 (Morganti et al.  2003a) and 4C~12.50 (see Fig.~) and the Seyfert
galaxy IC~5063 (Oosterloo et al.  2000).  Observations carried out recently
(using the WSRT) have revealed fast \HI\ outflows in a few more objects with
line widths ranging from 800 up to 2000 km/s.  The optical depths of the
broad, shallow absorption lines are low, typically around 0.001.  The
detection of these broad shallow absorption features represents a significant
challenge for present-day radio telescopes.  Broad bandwidths and a very
stable bandpass are needed.  Furthermore, we can currently only hope to detect
such weak features in very strong radio sources. Thus, the improvement that
the SKA will bring will be crucial to really understanding how common these
phenomena are, in which kind of sources they are observed, and what, indeed,
is their origin.

A number of possible hypotheses can be already made about the origin of the
gas outflow (e.g., starburst winds, radiation pressure from the AGN
adiabatically expanded broad emission line clouds, interaction of the radio
jet with the ISM).  However, to understand which of these physical process
drives the outflows we need to identify their exact location with high enough
resolution and sensitive observations, something quite difficult to carry out
with the available instruments.

\subsection{Jet masers}

The rich ISM in the central parsecs of AGN can also be detected via the
so-called ``jet masers''. Although H$_2$O megamasers are best known as a means
to probe the accretion disks, there is now evidence for a distinct class of
H$_2$O megamasers. In these sources, the amplified emission is the result of
an interaction between the radio jet and the encroaching molecular clouds.  The
only known sources in this class are NGC~1068 (Gallimore et al. 1996) and the
Circinus galaxy (Greenhill et al. 2001) which appear to have both
circumnuclear disk masers and maser emission arising along the edges of an
ionisation cone or outflow, and NGC~1052, in which the masers appear to arise
along the jet (Claussen et al. 1998).

One further example of this is Mrk~348, where the amplified emission is found
in VLBA studies to arise along the line of sight to a jet component and has a
very high linewidth (130 km s$^{-1}$), occurring on small spatial scales (0.25
pc) and varies on scales of days or months (Peck et al. 2003).  The
combination of these points suggest that the H$_2$O emission is more likely to
arise from a shocked region at the interface between the energetic jet
material and the molecular gas in the cloud where the jet is boring through,
than simply as the result of amplification by molecular clouds along the line
of sight to the continuum jet.  The orientation of the radio jets close to the
plane of the sky also results in shocks with the preferred orientation for
strong masers from our vantage point.  This hypothesis is supported by the
spectral evolution of the continuum source, which showed an inverted radio
spectrum with a peak at 22~GHz, later shifting to lower frequencies.  The
maser emission in this source allows us to determine the physical and chemical
conditions in the pre-shock gas very accurately.

A stronger ``jet maser'' has been found in NGC~1052 (Claussen et al
1998), but in general, these sources tend to be orders of magnitude
weaker than the better known ``accretion disk'' megamasers.  Recent
surveys have found that masers with peak flux densities of tens of mJy
have been found in 29\% of galaxies which have either some evidence of
interaction between the NLR and the jet, or have their radio jet axis
oriented close to the plane of the host galaxy (Peck et al 2004).
Even this high detection rate is probably limited only by the
sensitivity of current survey instruments, there may be many more such
sources.  This class of masers is distinguished by the broad shallow
shape of the line, rather than by the physical location, since these
very broad lines from very compact sources must be occurring in a
region as energetic as the central few pc of an AGN.  Thus, high angular
resolution is less of an imperative than high sensitivity, broad
bandwidth and bandpass stability, so the SKA would be able to detect
and monitor these sources where current instruments cannot.

\begin{figure*}  
\centerline{
\psfig{figure=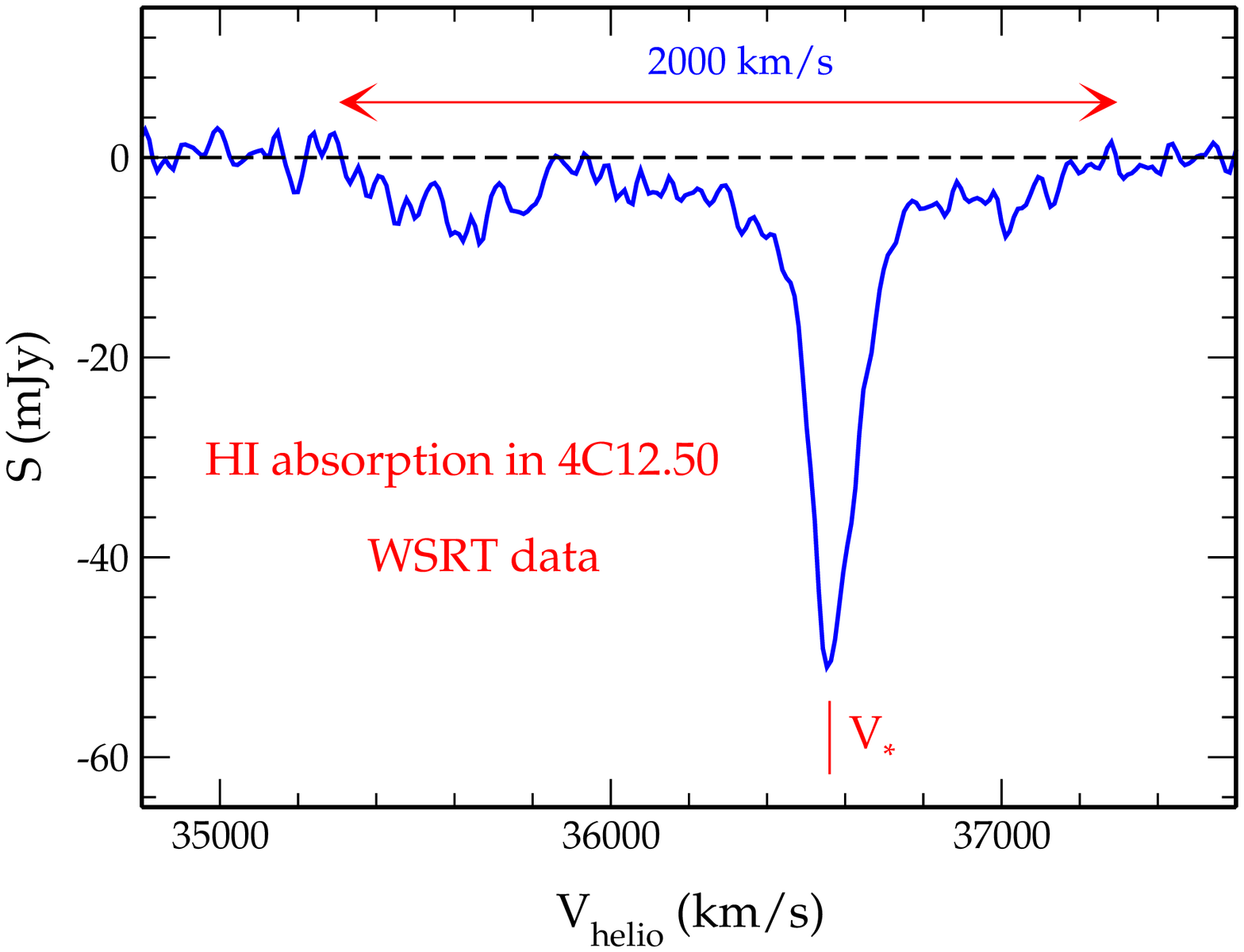,angle=0,width=7cm}   
\psfig{figure=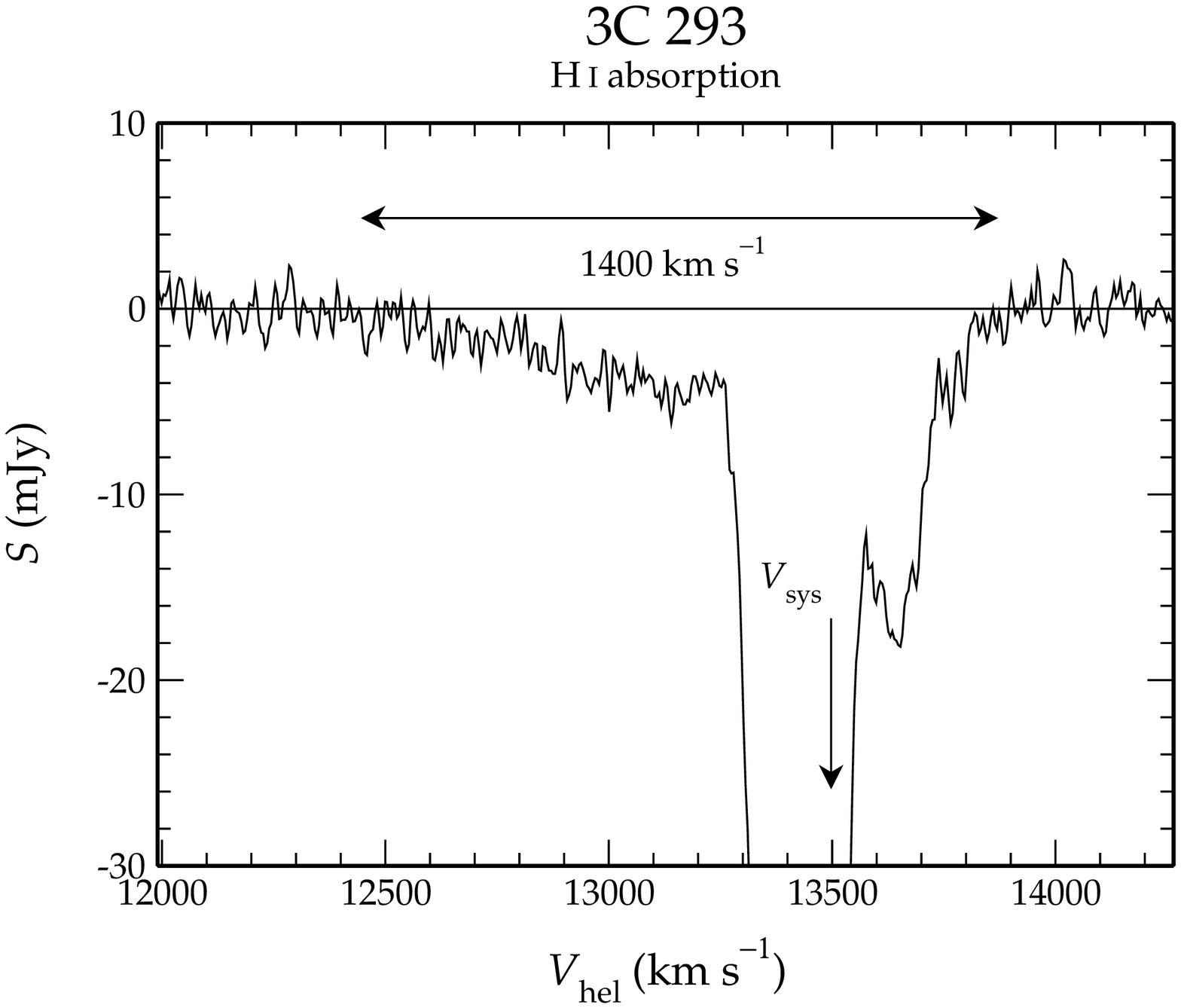,angle=0,width=7cm}}
\caption{The \HI\
absorption profile detected in 4C~12.50 (left) and 3C~293 (right) from the
WSRT observations (Morganti et al. 2003a,b).  The spectra are plotted in flux
(mJy) against optical heliocentric velocity in \kms.}

\end{figure*}

\section{The impact of SKA}

\subsection{ H$_2$O maser emission}

The SKA will contribute substantively to the detection and mapping of
H$_2$O maser sources in AGN, and perhaps to the detection of obscured
and distant low luminosity AGN through detection of their maser
emission (which can be strong even in ``weak'' AGN).  The typical line
strength of new detections has dropped by about an order of magnitude
per decade, to 0.01-0.05~Jy today, with a maximum distance of $\sim
230$~Mpc (see Figure 3, Greenhill this Volume). The rapid increase in source counts over the last few years
is attributable to the advent of observing systems with
higher sensitivity {\it and} instantaneous bandwidth at Effelsberg,
Green Bank, and Tidbinbilla.  Though the sensitivity of some existing
assets can be improved somewhat, order of magnitude improvements will
require use of the SKA or prototypes.
For line detection experiments, the SKA will be on the order of
$100\times$ more sensitive than the best observing systems today.
Instantaneous bandwidths on the order of 500~MHz per polarisation and
baseline will accommodate even the broadest maser source known today
(200~MHz) and provide a margin for detection of broader systems.
Observed line widths can be on the order of 1~km\,s$^{-1}$.  Hence,
the SKA backend channel spacing must be $< 30$~kHz to properly sample the
spectra (i.e., 16~K channels per baseline per polarisation).  The use
of an interferometer for line detection experiments promises better
(flatter) spectral baselines than can be achieved consistently over
very broad bandwidths and in the presence of background continuum
emission with most single dish instruments.  Perhaps more importantly,
deep, high frequency continuum mapping programs could be conducted in
parallel with maser surveys.  The simplest survey mode would entail
use of the SKA core so that the beam width is at least as large as the
galactic nuclei. However, if real-time imaging pipelines are
available, the full array could be used to search spectra constructed
from image cubes.

\begin{table*}

\caption{Approximate Array Sensitivities}

\begin{center}

\begin{tabular}{l|c|cc|c}
\hline
Array$^{(a)}$ & Relative    & \multicolumn{2}{|c|}{Self-cal}    & Posn-Swtch    \\
              & Sensitivity &${F_\nu^{peak}}^{(b)}$ & Synthesis$^{(c)}$
& Synthesis$^{(c)}$ \\
              &             &(mJy)                & (mJy hr$^{1\over2}$) & (mJy hr$^{1\over2}$)\\ 
\hline
\hline
VLBA                    & 1   & $\sim 1$   & 33  & 59       \\
VLBA+G                  & 3.1 & $\sim 0.2$ & 11  & 20       \\
VLBA+G+Y                & 6.4 & $\sim 0.2$ & 5.1 & 9.1      \\
VLBA+G+Y+E+$2\times$D   & 10  & $\sim 0.2$ & 3.2 & 5.7      \\
G+Y+E+$2\times$D        & 9.0 & $\sim 0.06$ & 3.7 & 6.6     \\
\hline
SKA core + $10\times 25$m  & 44  & $\sim 0.02$ & 0.74 & 1.3 \\
SKA core + $400\times 12$m & 100 & $\sim 0.05$ & 0.33 & 0.59\\
\hline
\hline
\end{tabular}
\end{center}

\vspace{-0.1in}
{ $^{(a)}$ G: GBT; Y: VLA; E: Effelsberg; D: DSN 70-m (Spain and US);
SKA core: proposed 5 km core of the Square Kilometer Array,
corresponding to 50\% of the total collecting area and treated as a
phased aperture.  Because of required high angular resolution, {\it
proposed 12-m and 25-m supplements to the SKA core must be external to
the SKA} in light of a planned 3000~km maximum extent.

$^{(b)}$ Minimum peak line strength for which self-calibration is
possible.  The adopted minimum requirement is a $3\sigma$ detection of
a reference Doppler component on individual baselines, assuming a
1~km\,s$^{-1}$ spectral channel, dual polarisation, 30 second
coherence time, and 10\% processing loss. Nominal system equivalent
flux densities (SEFD) are 1100~Jy (VLBA), 30~Jy (GBT), 40~Jy (VLA),
100~Jy (Effelsberg), 80~Jy (DSN-Spain), 90~Jy (DSN-US), and 0.56~Jy
(SKA core) for elevations of $\sim 40^\circ$. The SKA SEFD corresponds
to a high-frequency array specification of $A_e/T = 10^4$, where $A_e$
is effective collecting area and $T$ is system temperature.  For
arrays of large and small apertures, self-calibration requires the
$3\sigma$ detection to be achieved on baselines that link the large
and small apertures.

$^{(c)}$ Detection limit ($5\sigma$) for a 1 hour synthesis,
1~km\,s$^{-1}$ spectral channel, dual polarisation, 10\% processing
loss, and 80\% duty cycle on source for self-calibration or 25\% duty
cycle for position-switching with a nearby calibrator.  In arrays that
include the VLBA, antennas outside the continental US are assumed to
have half-tracks because of source visibility.
}
\end{table*}

The two most important ramifications of high SKA sensitivity for AGN
surveys will be completeness for nearby targets and extension to
distances $10\times$ greater than can be attempted with existing
instruments. It is difficult to predict how many sources will be
detected because statistics for present-day surveys remain poorly
understood.  However, in principal, the source sample will increase
three orders of magnitude, contingent in part on spectroscopic
identification of AGN at X-ray, optical, and near-infrared
wavebands. If the generic detection rate obtained in past surveys is
applied, and if the H$_2$O luminosity function is taken into account,
$10^5$-$10^6$ maser sources could be detected.

Interferometric mapping of sources that exhibit broadband emission
will continue to be a high scientific priority.  Self-calibration is
the most robust supporting technique, where a single Doppler component
is used to solve for complex station gains, which are subsequently
applied to other components.  This approach is feasible if there is a
line strong enough to be detected on individual baselines in an
atmospheric coherence time, or $\sim 30$ seconds at 22~GHz
(conservatively).  The phasing sub-arrays (or stations) within the SKA
would be critical for self-calibration to be practicable.  Weak maser
sources can be mapped with high fidelity if they lie close on the sky
to continuum calibration sources ($< 1^\circ$) and rapid position
switching of the array is possible. (Most maser sources do not lie in
nuclei that are also strong continuum emitters and would permit
continuum self-calibration at 22~GHz or below.)  For many maser
sources, the greatest limitations are presented by the absence of a
nearby, detectable calibrator with a well known position.  This
reinforces the importance of broad continuum observing bandwidths and
the inclusion of astrometry (i.e., enrichment of the celestial
reference frame) in the SKA mission.  Sensitivity limits for
self-calibration and position-switched observations are presented in
Table~1.

Although the SKA will substantially increase the number of known maser
sources, most key science (e.g., apart from statistical studies of detection
rates for different samples) will require high angular resolution imaging. For
a nominal maximum baseline of 3000~km, a $\sim 1$~pc radius disk inner edge
(e.g., NGC\,1068) will subtend the synthesised beam at a distance of only
$\sim 500$~Mpc, and detection of smaller disks (e.g., NGC\,4258) at such large
distances would be problematic.  If masers detected in each back-end spectral
channel are modelled as point sources, angular resolution limitations may be
offset via centroid fitting, which amounts to subdividing the beam in
proportion to the signal-to-noise ratio (the higher the $S/N$, the smaller the
uncertainties in the centroid position).  However, reliance on this technique
at a greater level than today, when intercontinental baselines are used for
mapping, forecloses the possibility of unambiguously detecting fine-scale
substructure (e.g., disk thickness, spiral waves) as well as robust
measurement of proper motions.

The simplest solution, contingent on geography, will be operation of outrigger
stations in combination with the core of the SKA, which itself would be best
operated as a phased element.  This could provide an order of magnitude
increase in sensitivity over present day interferometers (Table~1), depending
on the size of the outrigger array.  In gross terms, this would increase the
distance at which imaging is possible by half an order of magnitude and the
sample of targets by on the order of $30\times$, to a few hundred.  However,
even with intercontinental baselines, resolution of accretion disks such as
the one in NGC\,4258 will be difficult beyond on the order of 100~Mpc.  In
that sense, the SKA is a critical instrument, but it is just half of what is
needed for future robust investigation of accretion onto supermassive black
holes. The other half is space interferometry (i.e. the iARISE and VSOP2
missions operating in conjunction with the SKA).

\subsection{\HI\ 21cm absorption studies and SKA}

The study of neutral hydrogen in the surroundings of active nuclei is
limited, at present, to AGN with a relatively strong radio core. This
limit is even stronger if we are looking for weak and broad features,
associated with \HI\ with extreme kinematics.  These limitations will be
completely overcome by the SKA and this has a number of interesting
implications.  The most obvious is that for an improvement in
sensitivity of two orders of magnitude (combined with higher image
fidelity), we will be able to expand the study to radio-loud AGN with weak
cores, like most of the classical powerful radio galaxies. In this way, we
will expand our knowledge base on the occurrence of \HI\ associated
with circumnuclear tori and their structure.
\HI\ absorption with optical depths of the order of $\tau \sim 0.01$,
typical for the absorption that we consider is associated with the tori, will
then be detectable in sources as weak as a few mJy.

From this improvement, we will not only obtain a complete picture of the
presence of \HI\ in the nuclei of radio loud AGN, but we will also explore
the uncharted region of low luminosity AGN. The comparison of the structure of
the circumnuclear tori (as derived from the \HI) in low and high
(radio) luminosity AGN will tell us whether the differences in radio
properties are related to the way the accretion (low $vs$ high
rate/efficiency) occurs.

The new possibilities offered by the SKA will be even more crucial for the
detection and study of the shallow, broad ($\sim 1000$ \kms) \HI\ absorption.
The typical optical depth found so far for these components is $\tau \sim
0.001 - 0.0005$.  These values imply that, at present, we can only search for
such \HI\ with extreme kinematics in very strong radio sources (around 1~Jy or
more). Reaching this result has already required an effort in terms of
bandpass stability of present day telescopes and this issue will have to be
considered for the SKA as well.  However, even for spectral dynamic range of
the order of what can be reached today ($\sim 10^{-4}$), the SKA will allow us
to explore the presence of extreme kinematics in radio sources as weak as 10
mJy.  Moreover, depending on the lower frequency reachable by SKA, we will be
able to investigate the presence of low opacity, broad \HI\ absorption and in
particular the occurrence of outflows of neutral hydrogen, in high-$z$
sources, i.e. sources that are believed to be surrounded by a denser and
richer ISM.

Interestingly, for very strong radio sources the sensitivity of the SKA will
permit the study of gas at extremely low optical depth (of the order of $\tau
\sim 0.00001$). Thus, we will be able to explore the presence of gas in
different physical conditions (gas with very low column density or gas in
regions with high temperature) and extend the parameter space for exploring
the condition of the gas in the nuclear regions.  All these studies will also
require high spatial resolution (sub-arcsec, but possibly milliarcsec) that
will be essential for localising the absorption and understand its origin.

However, the SKA will go much further. It will also allow us to carry out
blind searches, a technique that we are just beginning to explore with the
present-day telescopes.  The spectral-line mode, in which  continuum
observations will be carried out as well, with its large instantaneous
bandwidth and large number of channels, facilitates the
serendipitous detection of \HI\ (in emission or absorption) from galaxies
in any observed field and in the redshift range of the observed band.  This
can be already done to some extent using WSRT standard continuum observations
but the SKA will allow us to investigate an unexplored region of
parameter space.  In the case of the WSRT deep survey of the Spitzer Space
Telescope First-Look Deep region (Morganti et al.  2004b), \HI\ emission was
detected in four galaxies in the central region of the field.  The sensitivity
of these observations (and the limited band), however, do not allow the
detection of \HI\ in absorption (except with extremely high optical
depths). All this will be routinely possible with the SKA and, with the
sensitivity expected, we could search for
\HI\ absorption at $\tau \sim 0.01$ level (the typical absorption found in
cases of circumnuclear tori) on every source in the observed field stronger
than only a few mJy. It will be like searching every source of the NVSS
catalogue for \HI\ absorption. The large instantaneous bandwidth will ensure
that a large range in redshift is covered to detect this absorption.

\section{Conclusions}

The dramatic improvement in spectral line sensitivity made possible by
the large collecting area of the SKA will benefit all types of  emission
and absorption  studies.  A much larger sample of host and background
sources will be available, and consequently we will be able to compare the
nuclear environments, as determined by these observations, over a wide
range of AGN luminosities, orientations, and evolutionary stages.  This is
essential to allow the effects of orientation, for example, to be disentangled
from other source properties.

The critical SKA parameters for these studies are high angular resolution at
moderately low frequencies (1-2 GHz) for imaging of HI absorption, OH masers,
and free-free absorption in AGN, and high angular resolution at high
frequencies (up to 22 GHz) for water masers.  In all cases, of course,
collecting area is the key to sensitivity.  However, very wide bandwidths and
high spectral dynamic range will also be  important parameters of the
SKA. All this will contribute to a significant and necessary step forward
in the understanding of the environment of AGNs.


\begin{thebibliography}{9}
\bibitem{} Antonucci R. 1993, ARA\&A 31, 473 
\bibitem{} Ball, G.  H., Greenhill, L.  J., Moran, J.  M., Henkel, C., \& Zaw,
I.  2004, ApJ, submitted

\bibitem{} Bicknell G., Saxton, C. J., Sutherland, R. S., Midgley, S.,
Wagner, S. J.  2003, New Astronomy Reviews, Volume 47, Issue 6-7, p. 537-544

\bibitem{}Braatz, J. A., Wilson, A. S., \& Henkel, C.  1996, ApJS, 106, 51

\bibitem{} Claussen M.J., Diamond P.J., Braatz J.A. et al. 1998, ApJL 500,
129

\bibitem{}Conway J.E.  1997, in {\it 2nd Workshop on GPS and CSS
Radio Sources}, Snellen et al.  M.N.  Publ JIVE, Leiden p.198

\bibitem{}Conway J.E.  1999 in {\sl Highly redshifted radio lines},
Carilli C.L. et al.  eds., ASP Conf.  Series 156, 259

\bibitem{}Conway J.E. \& Schilizzi R.T. 2000, in {\sl 5th European VLBI
Network Symposium}, Eds.: J.E. Conway et al., Onsala Space Observatory, p. 123

\bibitem{}Ferrarese, L., \& Merritt, D. 2000, ApJ, 539, L9

\bibitem{} Gallimore J.F., Baum S.A., O'Dea C.P. et al. 1996 ApJ 462, 740

\bibitem{}Goodman, J. 2003, MNRAS, 339, 937

\bibitem{}Greenhill, L. J. 2002, in ``Cosmic Masers: From Proto-Stars to 
Black Holes,'' eds. V. Migenes \& M. Reid, (San Francisco: ASP), 381

\bibitem{}Greenhill,  L. J. 2004, in ``Future Directions in High Resolution 
Astronomy: The 10th Anniversary of the VLBA,'' eds. J. D. Romney \&
M. J. Reid, (San Francisco: ASP), in press

\bibitem{}Greenhill, L. J., Gwinn, C. R. 1997, Ap\&SS, 248, 261

\bibitem{}Greenhill, L. J., Herrnstein, J. R., Moran, J. M., et al. 1997, ApJ, 486, L15

\bibitem{} Greenhill L.J., Moran J.M., Booth R.S. et al. 2001, in IAU
Sym. 205, {\it Galaxies and their Constituents at the Highest Angular
Resolutions}, R. T. Schilizzi ed., 2001, p. 334-337.

\bibitem{}Greenhill, L. J., Booth, R. S., Ellingsen, S. P., et al. 2003, ApJ, 590, 162

\bibitem{}Hagiwara, Y., Diamond, P. J., Miyoshi, M, et al. 2003, MNRAS, 344, L53

\bibitem{}Hagiwara, Y., Kohno K., Kawabe R., Nakai N. 1997 PASJ 49, 171

\bibitem{}Henkel, C., Braatz, J. A., Greenhill, L. J., et al. 2002, A\&A, 394, L23
\bibitem{}Henkel, C., Braatz, J.A., Tarchi, A. et al. 2004, in
{\it Dense Molecular Gas around Protostars and in Galactic Nuclei},
Y. Hagiwara, W.A. Baan \& H.J. van Langevelde eds.
 Ap\& SS  in press (astro-ph/0407161)

\bibitem{}Herrnstein, J. R., Moran, J. M., Greenhill, L. J., et al. 1997, ApJ, 475, L11

\bibitem{}Herrnstein, J. R., Moran, J. M., Greenhill, L. J., et al. 1999,
Nature, 400, 539

\bibitem{}Ishihara, Y., Nakai, N., Iyomoto, N., et al. 2001, PASJ, 53, 215

\bibitem{}Jones, D.L., Wehrle, A.E., Piner, B.G., and Meier, D.L., 2001, ApJ, 553, 968

\bibitem{} Koekemoer A. M., Henkel C., Greenhill L. J. Dey, A., van
Breugel W., Codella C., Antonucci R. 1995, Nature 378, 697

\bibitem{}Kondratko, P. T., Greenhill, L. J., \& Moran, J. M. 2004, 
ApJ, submitted (astro-ph/0408549)

\bibitem{}Levin, Y., \& Beloborodov, A. M. 2003, ApJ, 590, L33

\bibitem{}Maloney, P. R. 2002, PASA, 19, 401

\bibitem{}Maness H.L., Taylor G.B., Zavala R.T., Peck A.B. \& Pollack
L.K. 2004 ApJ 602, 123
\bibitem{} Mihos J.C. \& Hernquist L. 1996 ApJ, 464, 641

\bibitem{} Milosavljevi\'c, M., \& Loeb, A. 2004, ApJ, 604, L45

\bibitem{}Moran, J. M., Greenhill, L. J., \& Herrnstein, J. R. 1999, JApA,
20, 165

\bibitem{}Morganti R., Oosterloo T.A., Emonts B.H.C., van der Hulst
J.M., Tadhunter C.N. 2003a, ApJLetter 593, L69

\bibitem{}Morganti et al. 2003b, in {\sl Recycling Intergalactic and
Interstellar Matter}, IAU Symposium 217,  eds. P.-A. Duc et al.,
ASP, p. 332 (astro-ph/0310629)

\bibitem{}Morganti R. et al. 2004a A\&A 424, 119

\bibitem{}Morganti R., Garrett M., Chapman S., Baan W., Helou G., Soifer T. 
2004b A\&A 424, 371

\bibitem{}Mundell, C. G., Pedlar, A., Baum, S. A., O'Dea, C. P., Gallimore,
J. F. and Brinks, E. 1995, MNRAS, 272, 355

\bibitem{}Neufeld, D. A., \& Maloney, P. R. 1995, ApJ, 447, L17

\bibitem{}Oosterloo T.A. et al. 2000, AJ 119, 2085

\bibitem{}Peck A.B., Henkel, C., Ulvestad, J. S. et al. 2003 ApJ, 590, 149

\bibitem{}Peck A.B., Taylor G.B.  2001, ApJ 554, L147

\bibitem{} Peck A.B., Taylor G.B., Conway J.E. 1999 ApJ 521, 103

\bibitem{} Peck A.B., Taylor G.B. 1998, ApJ, 502, L23

\bibitem{} Pringle, J. E. 1996, MNRAS, 281, 357

\bibitem{}Readhead, A. C. S., Taylor, G. B, Pearson, T. J., \&
Wilkinson, P. N. 1996, ApJ, 460, 634

\bibitem{} Shakura, N. I. and Sunyaev, R. A. 1973, A\&A, 24, 337

\bibitem{}Silk J., Rees M.J. 1998, A\&A 331, L1

\bibitem{} Tarchi, A., Henkel, C., Chiaberge, M., et al. 2003, A\&A, 407, L33

\bibitem{} Taylor G.B et al. 2002 ApJ 574, 88

\bibitem{} Taylor G.B., O'Dea C.P., Peck A.B., Koekemoer A.M., 1999,
ApJ 512, 27

\bibitem{} Taylor G.B. 1996, ApJ 470, 394

\bibitem{}Trotter, A. S., Greenhill, L. J., Moran, J. M., et al. 1998, ApJ, 495, 740

\bibitem{}van Langevelde H.J., Pihlstr\"om Y.M., Conway J.E., Jaffe W.,
Schilizzi R.T. 2000, A\&A 354, 45

\bibitem{}Vermeulen et al. 2003, A\&A 404, 861

\bibitem{}Walker, R.C., Dhawan, V., Romney, J.D., Kellermann, K.I., and
Vermeulen, R.C., 2000, ApJ, 530, 233

\bibitem{}Watson, W. D. 2002, in {\it ``Cosmic Masers: From Proto-Stars to Black Holes''}, eds. V. Migenes \& M. Reid, (San Francisco: ASP), 464
\end{thebibliography}
\end{document}